# Complete polarization of electronic spins in OLEDs

T. Scharff, W. Ratzke, J. Zipfel, P. Klemm, S. Bange, and J. M. Lupton[*)]

Institut für Experimentelle und Angewandte Physik, Universität Regensburg,

Universitätsstraße 31, 93053 Regensburg, Germany

**At low temperatures and high magnetic fields, electron and hole spins in an organic light-emitting diode (OLED) become polarized so that recombination preferentially forms molecular triplet excited-state species. For low device currents, magnetoelectroluminescence (MEL) perfectly follows Boltzmann activation, implying a virtually complete polarization outcome. As the current increases, the MEL effect is reduced because spin polarization is suppressed by the reduction in carrier residence time within the device. Under these conditions, an additional field-dependent process affecting the spin-dependent recombination emerges, which appears to relate to the build-up of triplet excitons and the interaction with free charge carriers. Suppression of the EL at high fields on its own does not, strictly, prove electronic spin polarization. We therefore probe changes in the spin statistics of recombination directly in a dual singlet-triplet emitting OLED, which shows a concomitant rise in phosphorescence intensity as fluorescence is suppressed. Finite spin-orbit coupling in these materials gives rise to a microscopic distribution in effective g factors of electrons and holes, Δg, i.e. a distribution in Larmor frequencies, leading to singlet-triplet mixing within the electron-hole pair as a function of applied field. This Δg effect in the pair further suppresses singlet-exciton formation in addition to thermal spin polarization of the individual**

---

[*)] Corresponding author. Email: john.lupton@ur.de

**carriers. Since the $\Delta g$ process involves weakly bound carrier pairs rather than free electrons and holes, in contrast to thermal spin polarization, the effect does not depend significantly on current or temperature.**

A century ago, the experiment of Stern and Gerlach arguably led to one of the most important, exciting and unanticipated discoveries of physics.[1] While trying to measure the magnetic moment of atoms associated with the orbital angular momentum of electrons, the intrinsic magnetic moment of the electron was inadvertently discovered – the electron "spin", a property, which Dirac subsequently proved to be a direct consequence of the relativistic formulation of Schrödinger's matter-wave equation. Manipulation of the electron spin as inspired by the breakthrough of Stern and Gerlach has long been commonplace in "spintronics" technology,[2] in magnetic-resonance imaging,[3] and in quantum-information science;[4] it is even used in high-energy physics in studies of spin-dependent scattering processes.[5] But the spin of an electron is also central to the operational principles of seemingly disparate processes: in an organic light-emitting diode (OLED), for example, electrons and holes recombine in a spin-dependent fashion to generate light. Since the recombinant singlet and triplet molecular excited-state species are non-degenerate, the statistical distribution of the spin permutation symmetries of the recombination products determines the overall power efficiency of a device.[6] Similar phenomenology appears at the heart of photochemical reactions triggered by photoinduced electron transfer.[7] It has, for example, been proposed that retinal pigment-protein complexes of some bird species support the formation of spatially separated spin-correlated carrier-pair states, the singlet-triplet recombination yield of which can be influenced by the changes in spin precession arising on magnetic-field scales as small as geomagnetic-field strengths.[8] These processes depend on the permutation symmetry of a spin pair, and not on a Stern-Gerlach type of spin polarization.

OLEDs offer a potentially interesting semiconductor basis to investigate spin polarization phenomena because spin-orbit coupling (SOC) effects are generally weak.[9-14] In terms of their interactions, electron and hole spins in an organic semiconductor effectively behave as those of free electrons in vacuum, with g factors of close to 2. OLED-like devices with ferromagnetic electrodes exhibit a range of intriguing magnetoresistance and magnetoelectroluminescence (MEL) phenomena, which suggest that it may be possible to inject spin-polarized currents into organic semiconductors.[15,16] However, it is very challenging to prove with certainty that a net magnetization is actually established within the organic semiconductor since it has not been possible to apply conventional techniques of spin spectroscopy such as Kerr- and Faraday-rotation.[17,18] Spin-polarized photoelectron spectroscopy may be sensitive to the interface of the ferromagnetic electrode and the semiconductor,[19] whereas muon scattering provides a somewhat indirect measure of electronic spin polarization and requires major research infrastructure.[20]

A conceivable way to examine the emergence of effective spin polarization is through the substantial change in spin statistics of the recombinant electron-hole pair, i.e. the change in the ratio of molecular singlet to triplet excitations. However, OLED structures with potentially spin-polarizing ferromagnetic electrodes show only minimal relative changes in intensity, of order of a few percent.[21,22] More importantly, most OLED materials emit only from either the singlet or the triplet state, but not from both. It is therefore not possible to conclusively assign the loss or gain in intensity within one recombination channel with certainty to a change in spin statistics alone. The most compelling demonstration to date of spin polarization occurring in an OLED structure reported a suppression of EL intensity by almost 50 % due to thermal spin polarization (TSP).[23] When the Zeeman energy of a spin in a magnetic field oriented along the $z$-axis becomes comparable to the thermal energy $k_B T$, the product of Boltzmann's constant and temperature, the probability of spins polarized along $\pm z$ at thermal equilibrium becomes a function of temperature and magnetic field,

$$P_\uparrow^{\text{eq}} = \frac{1}{1+\exp\left(\frac{g\mu_{\text{B}}B}{k_{\text{B}}T}\right)} \quad \text{and} \quad P_\downarrow^{\text{eq}} = 1 - P_\uparrow^{\text{eq}}\,. \tag{1}$$

Unpolarized injected spins equilibrate due to the spin-lattice relaxation within a time $\tau_s$ to a distribution given by eq. (1). However, the lifetime of the free spins in the organic semiconductor is limited by the formation of the strongly bound excitons, so that the resulting spin-statistics is composition of unpolarised and polarized spins. The degree of mixture of both is then defined by the relaxation time $\tau_s$ and the time available for relaxation, the free-carrier lifetime $\tau_c$. Eq. (1) must therefore be modified to

$$P_\uparrow = \left(\frac{1}{2} - P_\uparrow^{\text{eq}}\right)\exp\left(-\frac{\tau_c}{\tau_s}\right) + P_\uparrow^{\text{eq}}, \tag{2}$$

Spins tend to align in parallel to the external magnetic field, so that recombination preferably arises in the triplet state. In a material which exhibits only fluorescence from the singlet and no phosphorescence from the triplet state, the overall EL yield, the probability of light emission from singlets, which drops with increasing magnetic field following Boltzmann activation,

$$P_S = P_\uparrow \cdot P_\downarrow = P_\uparrow - P_\uparrow^2\,. \tag{3}$$

While Wang *et al.* indeed showed a suppression of EL intensity with magnetic field,[23] the form of the functional field dependence left room for interpretation. More importantly, a suppression of EL intensity alone does not, strictly, prove spin polarization – to do this, an anticorrelation must be demonstrated between the reduction in the yield of singlets and the increase in triplet yield. Here, we demonstrate a complete suppression of the EL by strong magnetic fields at low temperatures in a fluorescent OLED material. The degree of EL suppression depends on temperature, but also on current, with the effect of the apparent spin polarization vanishing with increasing temperature and current. We introduce a dual singlet-triplet emitting compound,[24-26] which shows a direct anticorrelation between the suppression of singlet fluorescence and the increase of triplet phosphorescence due to the formation of a spin-polarized ensemble of electron-hole pairs. The approach offers a direct visualization of the Stern-Gerlach-type of spin polarization in OLEDs – a spectroscopic analogue of space quantization; and reveals subtle SOC effects in these low-atomic-order-number materials. TSP is shown to be superimposed by SOC-related spin-mixing effects, which are responsible for a microscopic distribution in effective Larmor frequencies of the spins, i.e. in effective g factors. When the effect of TSP is suppressed at high currents and temperatures, this $\Delta g$ effect[27-34] becomes apparent as a further mixing channel between singlet- and triplet-type spin pairs. OLEDs therefore offer an approach to differentiate between changes in spin-dependent recombination due to single-carrier spin polarization effects[35] and spin-pair correlation effects.[36]

Although there are several reports of OLED magnetoresistance[31] and photoconductivity[37] at high magnetic fields, given the experimental challenges involved, there are only two prior studies of MEL under such conditions.[23,38] To accurately measure the EL intensity at very low currents and brightnesses requires the use of a single-photon counter which is not affected by stray magnetic fields of the cryostat. Photomultipliers and other optical intensifiers cannot be used for this challenge as the Lorentz force exerted by the stray fields induces substantial artefacts in the detector sensitivity. Point detectors such as avalanche photodiodes are also of limited use since the OLED emits over a surface area and any elastic deformation of the device, sample holder or cryostat due to magnetic forces will induce artefacts in the measurement which cannot be accounted for. Instead, we use a single-photon counting area detector, a scientific CMOS (sCMOS) camera,[26] which offers a detection quantum efficiency of almost 80 % together with the ability to correct for any unwanted mechanical movement, provided that the image of the OLED pixel projected onto the camera is smaller than the area of the camera chip. Figure 1 shows the change in EL intensity of an OLED made of the commercial poly(phenylene-vinylene) (PPV) derivative "super-yellow PPV" (SYPPV), operated in constant-current mode, at a temperature of 1.5 K and a very low current of 550 nA, corresponding to a voltage of around 5.5 V. The sketch illustrates how TSP arises as the Zeeman splitting of spin states becomes comparable to the thermal energy $k_BT$. The entire measurement takes 4 hours for two complete sweeps up and down in magnetic field, and the curve shows the average of both sweep directions. The functionality perfectly follows the prediction of equation (3), with the only free fittin parameter being the ratio of spin residence time in the field to spin-lattice relaxation time, $\tau_c/\tau_s = 5.63$. The fit suggests that a spin polarization of >99 % can be reached.

In contrast to the report of Wang *et al.*,[23] it was not possible to extract spin residence and relaxation times directly from transient EL measurements. Wang *et al.*[23] appeared to observe an additional slow relaxation process at high magnetic fields which they attributed to the TSP mechanism. However, the transient EL measured at low currents displayed in Supplementary Figure S1 shows clearly that the magnetic field actually quenches the fluorescence immediately, i.e. there are no resolvable EL dynamics associated with TSP; no additional relaxation processes appear at high fields. This absence of additional dynamics implies that the observed transient EL originates from spin-polarized charge carriers, i.e. the transient dynamics bear no relation to spin-lattice relaxation.

Next, in Figure 2a, we probe TSP by the temperature dependence of the OLED MEL. As expected, an increase of the temperature results in a weaker polarization of the spins and therefore in a reduced suppression of the EL at strong magnetic fields.[23] The black lines show the predicted functionalities from eq. (3), which are in excellent agreement with the measurements using one common $\tau_c/\tau_s$ value of 3.7 for all datasets. The reason these experiments have to be performed at such low currents is apparent from the current dependence of TSP shown in Figure 2b. As the current increases, the overall magnitude of the MEL effect is reduced,[23] suggesting that the TSP is quenched. For TSP to occur, carriers must reside in the device for a sufficiently long time to be able to equilibrate in the field by spin-lattice relaxation.[23] The higher the device current, the shorter the carrier residence time and the greater the probability of scattering between two carrier spins becomes. This latter effect will tend to randomize the spin orientations again within the device.[23] In addition, the shape of the functionality of the MEL curve changes with increasing current, as seen in the direct comparison of two curves in Figure 2c: MEL can no longer be described accurately by eq. (3). It is known that triplet excitons can interact with free charge carriers or with each other to annihilate and generate additional singlet states, thereby forming a singlet population which exceeds the spin-statistical limit.[39,40] An external magnetic field tends to inhibit these processes, thereby reducing the overall number of singlets formed besides the TSP singlet-suppression effect.[41]

To test for the possible influence of such delayed fluorescence from triplet-triplet annihilation, we examined time-resolved EL of the afterglow of the OLED, i.e. the turn-off behaviour, as shown in Supplementary Figure S2. We observe only a sharp EL overshoot at turn-off, followed by a weak afterglow over a few microseconds, which is unaffected by temperature, magnetic field and voltage – a clear indication of the release of trapped or accumulated charge carriers in the device.[42] In contrast, time-resolved measurements of the turn-on behaviour, shown in Supplementary Figure S3, indicate the emergence of an additional rise in EL intensity at high drive currents. In apparent disagreement with the earlier observations of Wang et al.,[23] these features are sensitive to temperature and magnetic field. However, as would be expected, the field dependence vanishes for temperatures above 50 K, indicating that TSP is only relevant at low temperatures where the Zeeman splitting energy exceeds the thermal energy. The sharp EL overshoot at turn-on suggests the accumulation of charge carriers at internal energy barriers.[43] From these observations, we conclude that there is no significant delayed fluorescence from triplet-triplet annihilation in our devices, but instead a substantial effect from accumulated or trapped charge carriers arises. These carriers of limited mobility can be fully polarized by the magnetic field and can also interact with the long-lived triplet excitonic states.[44] The pronounced change of MEL curvature at high currents, leading to a strong deviation from eq. (3), is therefore likely a signature of triplet-exciton polaron quenching[45] with accumulated spin-polarized polarons.

Suppression of the singlet yield alone does not imply the generation of spin polarized carriers in the OLED: conversion of singlet pairs to superposition-state triplet pairs suppresses fluorescence but does not lead to spin polarization. A spectroscopic test of signatures of the formation of polarized spins requires a comparison of the populations of singlet and triplet excitons and an examination of the functionalities of the dependence on magnetic-field strength. To do this, we introduce a surprisingly simple OLED emitter with intrinsic dual singlet-triplet emission because of the SOC arising from the mixing of $\pi$-orbitals with the non-bonding orbitals of a phenazine moiety.[24-26] The structure of 11,12-dimethyldibenzo[a, c]phenazine (DMDB-PZ) is shown in Figure 3a. The material is coevaporated with 4,40-

bis(N-carbazolyl)-1,10-biphenyl (CBP) to yield an OLED with distinct dual singlet and triplet emission. A representative EL spectrum, recorded at a temperature of 1.5 K, is plotted in Figure 3a. The inset shows a transient EL measurement, which demonstrates a triplet phosphorescence lifetime $\tau_{phos}$ of 170 ms at 1.5 K. Unlike conventional organometallic triplet emitters,[6,46] SOC is comparatively weak in these compounds, so that the singlet population is only partially quenched by intersystem crossing (ISC) to the triplet manifold. Because singlet and triplet are highly non-degenerate, the ratio of singlet to triplet emission intensity is not affected by temperature as in thermally activated delayed fluorescence.[47-49] Also, since every single chromophore shows dual emission, the detected singlet-triplet ratio in EL does not depend on triplet diffusion to an emitting site. Unlike dual-emitting conjugated polymers previously used to probe the evolution of the singlet-triplet ratio in MEL,[50] the ratio of fluorescence to phosphorescence is therefore not inherently temperature dependent.[51] To measure the spectrally resolved MEL effect of these dual-emitting OLEDs requires a substantial modification to the optical imaging setup. Filters are used to select the spectral regions marked in the spectrum in Figure 3a, allowing changes in EL intensity to be resolved down to levels of 1,000 ppm. Figure 3b shows the evolution of singlet and triplet EL intensity with magnetic field, at a temperature of 1.5 K and for a device current of 14 μA. The fluorescence is quenched by up to 98 %, but the maximal simultaneous increase in triplet intensity is only 34 %. There are two reasons for this. First, due to symmetry rules, only one of the three triplet states couples to the singlet manifold to give phosphorescence. Given the long phosphorescence lifetime of 170 ms of this material, spin-lattice relaxation can arise within the triplet excited state to mix the different triplet sublevels,[25] but not all triplets formed by spin polarization will contribute to phosphorescence.[25] Second, if we assume that, at low magnetic fields, the three triplet sublevels are populated equally by the recombination of electrically injected, unpolarized charge carriers, then the maximal change in the overall triplet population achievable by TSP of the carriers is 1/3 by transfer of all singlet pairs to the $T_+ = |\uparrow\uparrow\rangle$ state. It is therefore reasonable to observe an amplitude of TSP in the triplet phosphorescence channel corresponding to approximately 1/3 of the change seen in the singlet fluorescence channel.

Neither fluorescence nor phosphorescence can be described purely by the Boltzmann-type TSP functionality of eq. (3) (Figure 3b, left), indicating the presence of a further spin-mixing process in addition to TSP. Since spin-mixing processes related to hyperfine interactions, exchange interactions and the zero-field splitting occur at much smaller magnetic-field strengths, i.e. below hundreds of militesla,[37] we attribute this second mechanism to the "Δg effect" illustrated in Figure 3c. Because of the weak but finite SOC in these materials,[14] a distribution in effective g factors of electrons and holes exists within a carrier pair, implying that the carriers precess at slightly different frequencies in an external magnetic field. For small fields, spin precession is dominated by the local disorder in effective magnetic fields, which arises due to the distribution of nuclear magnetic moments, i.e. the hyperfine coupling.[52] At high fields, SOC effects dominate the precession frequency: spins in singlet carrier pairs $|\uparrow\downarrow\rangle - |\downarrow\uparrow\rangle$ precess to form superposition-state triplet pairs $|\uparrow\downarrow\rangle + |\downarrow\uparrow\rangle$.[29]

To a first approximation, the suppression of the yield of singlet excitons due to the Δg-mechanism in the high-field regime has been described by a phenomenological Lorentzian functionality,[37,53] which is rationalized simply in terms of the damping over time of the coherent spin mixing between singlet and triplet configurations of the carrier pair. Adding such a functionality to eq. (3) gives a phenomenological expression[31] of the form

$$P_{\mathrm{S}} = P_\uparrow - P_\uparrow^2 + \alpha \frac{\Delta B_{\frac{1}{2}}^2}{\Delta B_{\frac{1}{2}}^2 + B^2} \qquad (4)$$

for the singlet yield, which shows remarkably good agreement with the measurement on the righthand side of Figure 3b. The fitting parameters used involve only the ratio of spin residence to relaxation times, $\tau_c/\tau_s$, an amplitude $\alpha$ for the $\Delta g$ effect, and a width of the Lorentzian $\Delta B_{1/2}$, which relates to the difference in precession frequencies of electron and hole within the pair. The fitting procedure, described in the Supplementary Material, ignores data points below ±800 mT because the precession frequency of $\Delta g$-mixing is determined solely by SOC at high fields, whereas spin mixing due to the hyperfine fields dominates at low to intermediate fields.[53]

For the MEL in the phosphorescence channel, i.e. the OLED triplet formation yield, the modification in eq. (4) improves the fit compared to the pure TSP functionality of eq. (3), but slight discrepancies with regards to the experimental data remain. Considering the very long triplet excited-state lifetime and the possibility of generating additional triplet states due to TSP and the $\Delta g$ effect, a substantial accumulation of triplet excitations is expected to occur in a device. This accumulation will result in a higher quenching probability of triplets by either triplet-triplet annihilation or by triplet-polaron interactions[45] since the overall triplet population increases at large magnetic fields. These additional dynamic processes are not taken into account in this simple static model. We stress that the fit function is entirely qualitative in nature and cannot be used to extract effective SOC strengths of the material since the fit to the $\Delta g$ effect is not sufficiently sensitive to $\Delta B_{1/2}$ over the limited range of magnetic fields probed in the experiment.[31] It serves predominantly to extract the degree of TSP from the MEL. To determine $\Delta g$ precisely,[30] high-field electron paramagnetic-resonance spectroscopy is necessary, i.e. EPR measurements at static fields where hyperfine-induced broadening of the resonance no longer dominates the resonance spectrum as recently demonstrated in the context of OLEDs.[14,31,54]

Resolving this direct anticorrelation between the intensities of fluorescence and phosphorescence due to TSP requires careful optimization of the device structure and operating conditions to maximize the residence time of uncorrelated free charge carriers in the external magnetic field. In some device structures, the carriers may become trapped at the interface between two layers. While this trapping raises the effectiveness of TSP, it can also enable direct injection into the triplet state,[55] so that the ratio of singlet to triplet emission does not perfectly reflect the equilibrium spin statistics. However, changes in spin statistics will still be captured by changes in the fluorescence-to-phosphorescence ratio. Figure 4 illustrates how an increase in temperature or device current has a dramatic effect on the MEL functionality, which can appear to be quite distinct in the singlet and triplet channels. As in the case of SYPPV OLEDs (Figs. 1, 2), as the device current at a temperature of 1.5 K increases in Fig. 4a,b, the overall amplitude of the MEL effect decreases. The same trend is observed by increasing the temperature in Fig. 4 c,d. In both cases the shape of the curve appears to broaden because of the increased relative contribution of the $\Delta g$ effect to the MEL as indicated by the fit of eq. (4) to the data. These devices also show a sharp MEL feature around zero field, which presumably results from the spin-mixing process arising from electron and hole spin precession within the hyperfine fields and is not accounted for with the simple model used here.[50] Interestingly, this feature appears sharper and more pronounced in the increase of phosphorescence intensity with magnetic field than in the suppression of the fluorescence. This observation on the scale of several hundred mT is in contrast with measurements on similar diodes in the range of a few mT,[26] which shows a quantitative anticorrelation between singlet and triplet and the opposite effect with a rise in singlet yield and a suppression of triplets. At fixed temperatures, as the current is lowered, the hyperfine-field feature in the triplet channel appears to become sharper. As the current in panel b is raised to 100 μA, the functionality even inverts, so that the slope of the MEL of singlet and triplet have the same sign, i.e. the two intensities are no longer anticorrelated in contrast to the conventional situation.[26] This inversion indicates a sensitive interplay between effective carrier lifetime, which is related to the device current, and the hyperfine-induced spin mixing,[26] in addition to possible level-crossing effects in the triplet exciton at intermediate field strengths which complicates the MEL functionality substantially at fields below where TSP becomes significant.

TSP is a rather subtle effect, which is easily quenched under strong currents. However, the fact that TSP can be observed at near unity polarization implies that electron paramagnetic resonance signals should become huge at high fields and very low temperatures. In the dual emitters we expect this effect to become so strong that illumination of the device with THz radiation resonant with the magnetic-dipole transitions of the Zeeman-split spin states will visibly alter the singlet-triplet ratio and hence the emission colour of the OLED. Pulsed magnetic resonance experiments should then provide unique means to differentiate between coherent and incoherent spin-mixing processes, revealing how spins in organic semiconductors interact with each other[56] and with the environment. An intriguing question is whether the dramatic electronic spin polarization can be transferred to the nuclear spin bath as, for example, in silicon devices[57] or in anthracene crystals.[58] Hyperfine interactions may lead to hyperpolarization of the nuclei, which would alter the low-field magnetoresistance and MEL response. At present, there is no clear route to transferring the device swiftly from the high-field to the low-field regime given the inherent inertia in magnetic-field sweeps of superconducting magnets. Alternatively, nuclear magnetic resonance should also have an appreciable effect on EL under these conditions, providing complementary access to the hyperfine interaction to probe for nuclear spin polarization.[52] We expect that the conclusive observation of near-perfect spin polarization in OLEDs will offer further insight into the properties of electron spins in aromatic hydrocarbons, and may help explain the puzzling decoupling of spin and charge transport in these materials,[59] which appears to be responsible for some of the magnetoelectronic effects reported previously.

Funded by the Deutsche Forschungsgemeinschaft (DFG, German Research Foundation) – Project-ID 314695032 – SFB 1277.

## Methods

*OLED fabrication*

Glass substrates covered with 100 nm of indium-tin oxide (ITO) were obtained from Präzisions Glas & Optik (Germany) and structured by etching the ITO in $FeCl_3$/HCl solution. Next, the surface was treated in an ultrasonic cleaner in successive baths of acetone, 2 % Hellmanex III solution (Hellma Analytics), and isopropanol. After each cleaning step, the substrates were flushed with ultrapure water. Finally, they were treated by oxygen plasma cleaning (40 kHz, 80 W, plasma technology GmbH) for 30 minutes followed by UV/ozone exposure (Novascan Technologies) on a hot plate at 100 °C for a further 30 minutes. Directly after cleaning, an 80 nm thick hole-injector layer of poly(*3,4*-ethylenedioxythiophene) polystyrene sulphonate (PEDOT:PSS; Clevios P VP AI 4083, Heraeus) was spin coated on the ITO surface and the substrates were transferred to a nitrogen glovebox where they were baked out on a 150 °C hotplate for 30 minutes. For the emissive layer of the fluorescent SYPPV devices, 5 mg·ml[-1] of SYPPV was dissolved in toluene and spin coated to give a thickness of ~100 nm. For the top electrodes, barium (3 nm) and aluminium (250 nm) layers were deposited by thermal evaporation through a shadow mask, yielding an active pixel area of 3 mm². The dual-emitting devices were produced in a similar fashion except that the emitting layer was deposited by thermal evaporation. For the emissive layer, 4,40-bis (N-carbazolyl)-1,10-biphenyl (CBP, Ossila Ltd.) and 11,12-dimethyldibenzo[a,c]phenazine (DMDB-PZ, Sigma Aldrich) were deposited by thermal co-evaporation at a ratio of 97:3 to yield a thickness of 60 nm. All devices were encapsulated with a 500 nm thick layer of N,N′-bis(3-methylphenyl)-N,N′-diphenylbenzidine (TPD, Ossila Ltd.) by thermal evaporation to protect from air and to reduce thermal stress during cooling.

*Magnetic-field setup*

The samples were placed in a cryostat with a split-coil superconducting magnet and windows for optical access (American Magnetics). For the fluorescent devices, the EL emission was directly projected onto a sCMOS camera (Hamamatsu) using a lens system with the detector placed roughly 0.7 m away from the magnet. For the dual-emitting devices, an additional optical unit was positioned in the beam path, which allowed the spectral separation of fluorescence and phosphorescence. First, the incident light impinges on a dichroic mirror, which splits the beam at 532 nm. To achieve further spectral separation, a 500 nm shortpass filter and a 600 nm longpass filter were included in the beam paths. Both beams were focused into non-overlapping images onto the same sCMOS sensor, which allows a parallel and synchronous detection of fluorescence and phosphorescence. Alternatively, the EL emission could be focused onto a spectrometer. For detection, we used a gated iCCD camera (Andor iStar 720), allowing for time and spectrally resolved EL measurements. The OLEDs were operated in constant-current mode using a low-noise source-measure unit (Keithley 2400). During the measurement, the device voltage was recorded by an additional dc digital multimeter (Keysight 34470 A). The sweep rates of the magnet were 5 mT·s$^{-1}$ for the range 0 T<|B|≤6.4 T and 4 mT·s$^{-1}$ for 6.4 T<|B|≤8 T. The magnetic field was applied in a plane

perpendicular to the sample surface. The cryostat comprises a variable temperature insert

(VTI) which allows temperatures between 1.5 K and room temperature to be reached.

*Fitting procedures*

Details of the fitting parameters are given in the Supplementary Information Tables S1-S4.

**Data availability**

The raw data that support the plots within this paper and the other findings of this study are available from the corresponding author upon reasonable request.

**Figures and captions**

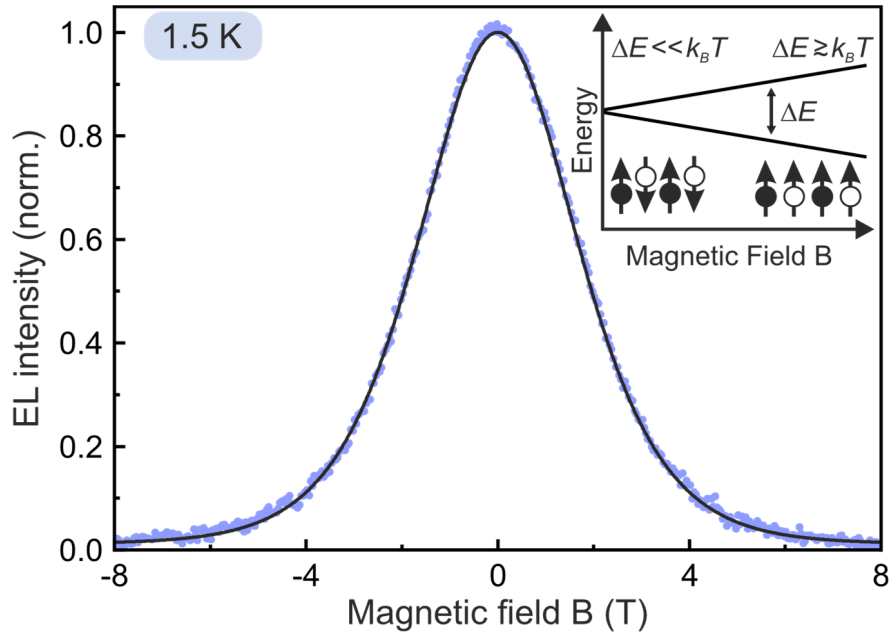

Figure 1. **Thermal spin polarization (TSP) in a "super-yellow" poly(phenylene-vinylene) (SYPPV) OLED at 1.5 K.** The magnetic field splits the energy of the spins with respect to their orientation in the field. Once this splitting exceeds the thermal energy $k_\mathrm{B}T$, spins can become polarized so that electron and hole pairs recombine preferentially into the triplet excited-state manifold of the molecule, which generally does not emit light. The electroluminescence (EL) intensity of the OLED is recorded as a function of magnetic-field strength. The change in EL intensity with magnetic field, the magnetoelectroluminescence (MEL), follows Boltzmann activation of the singlet-triplet ratio [black line, eq. (3)], demonstrating a degree of polarization of over 99 % at $\pm 8$ T.

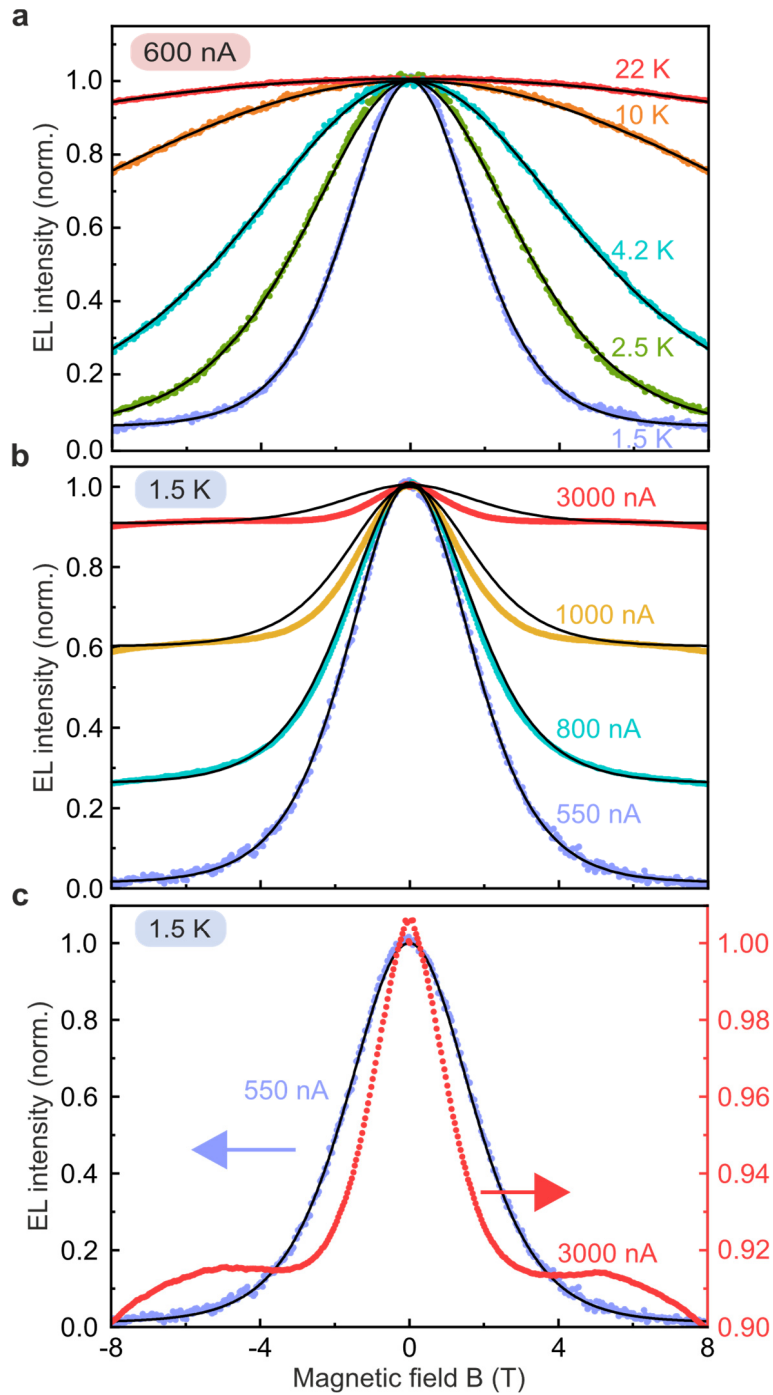

Figure 2. **Current and temperature dependence of TSP-induced MEL of a SYPPV OLED.** a) Temperature dependence at a device current of 600 nA. At elevated temperatures, the functionality of the MEL curves is in good agreement with the Boltzmann-type activation expected of TSP. b) Current dependence at a temperature of 1.5 K. At higher currents, the functionality narrows with respect to the Boltzmann activation behaviour [black lines,

following eq. (3)], most likely because long-lived triplet excitons are quenched by free polarons to form additional singlets. c) Direct comparison of MEL curves at 550 nA and 3000 nA drive current. Note the different scales for the blue and red curves.

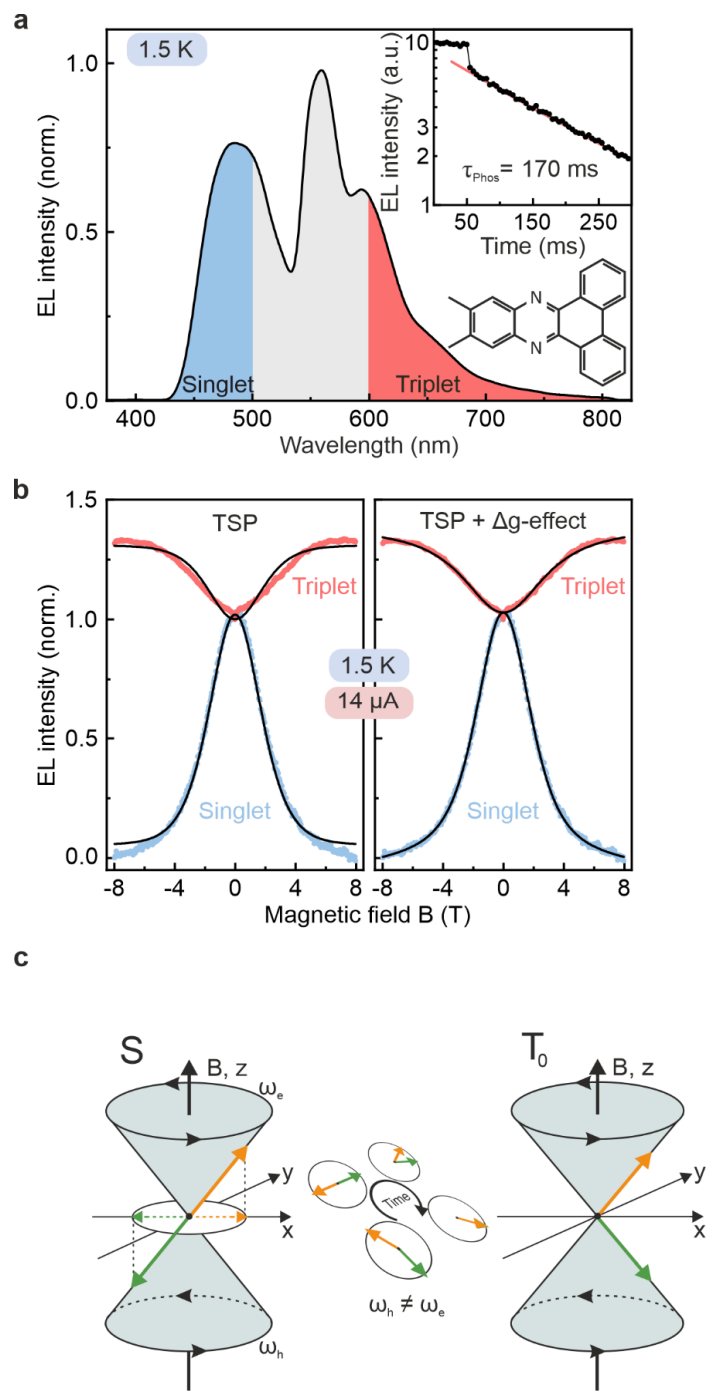

Figure 3. **MEL of dual singlet-triplet emitting OLEDs, resolving a suppression of singlet fluorescence with a concomitant rise in triplet phosphorescence.** a) EL spectrum of an OLED with the dual emitter DMDB co-evaporated with CBP, at 1.5 K. The shaded regions indicate the spectral range selected by filters to distinguish between fluorescence (blue) and phosphorescence (red). The inset shows a transient EL measurement, revealing a triplet

phosphorescence lifetime of 170 ms. b) Change of singlet and triplet luminescence intensity at 1.5 K under a device current of 14 μA, with a fit of the Boltzmann-TSP function [eq. (3)] and a fit of the TSP functionality modified for the Δg effect [eq. (4)]. c) Sketch of the singlet-triplet mixing process in the pair due to the Δg effect.

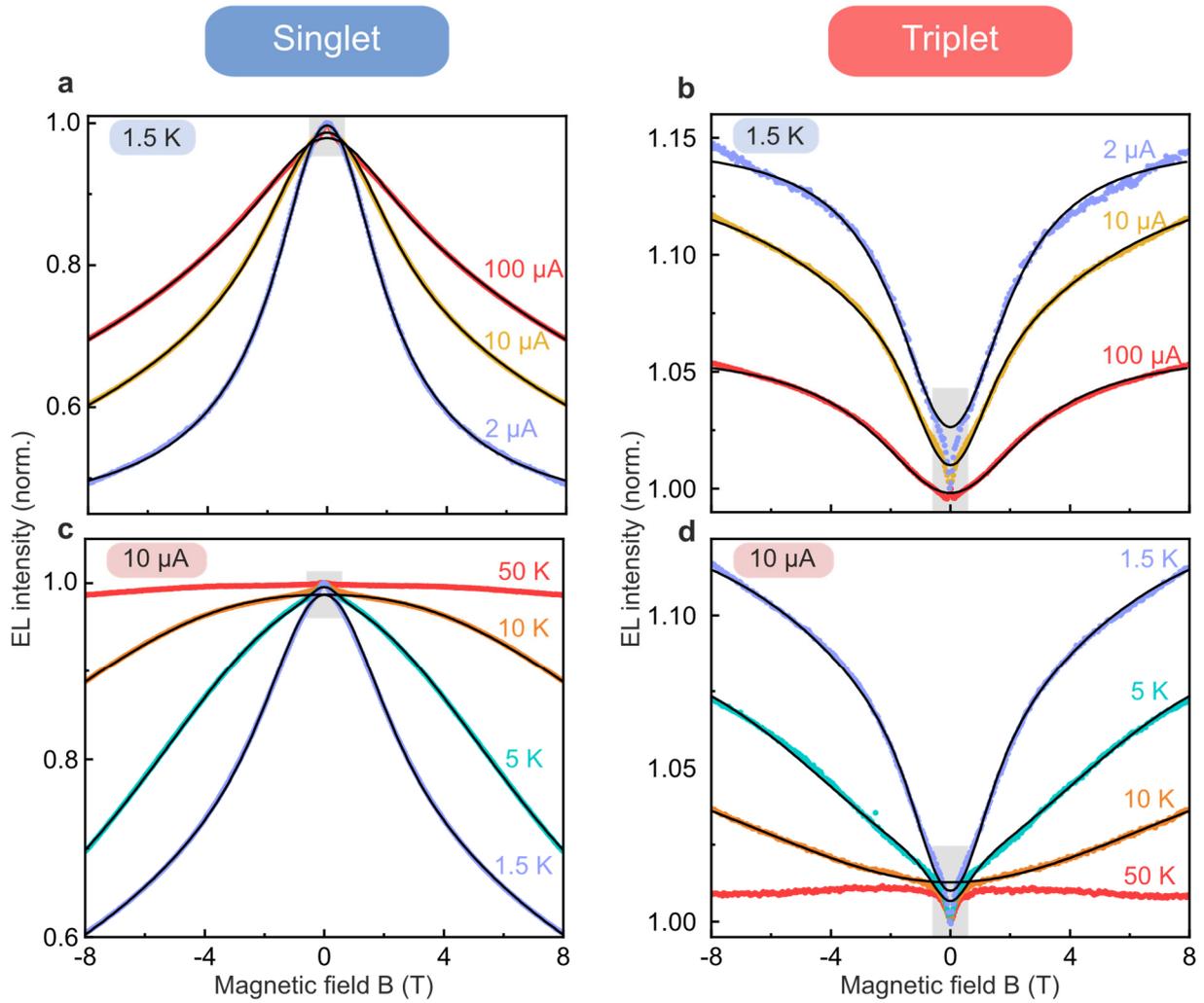

Figure 4. **Current and temperature dependence of the singlet and triplet EL intensity of a dual-emitting OLED.** The current dependence (a, b) is shown at a temperature of 1.5 K and the temperature dependence (c, d) at a current of 10 µA. The lines show a fit to the data of the combined functional dependence of TSP and the Δg effect following eq. (4). At higher currents and at low temperatures, a narrow feature is observed around zero field (grey areas), which is attributed to the effect of spin mixing due to hyperfine interactions. This grey region is excluded in the fitting procedure. The triplet intensity is quenched at high fields for the 50 K curves at 10 µA, so there is no meaningful fit possible.

**Supplementary Information**

**Complete polarization of electronic spins in OLEDs**

T. Scharff, W. Ratzke, J. Zipfel, P. Klemm, S. Bange, and J. M. Lupton

Institut für Experimentelle und Angewandte Physik, Universität Regensburg,

Universitätsstraße 31, 93053 Regensburg, Germany

**Time-resolved EL measurements**

Time-resolved EL was measured by applying voltage pulses of 20 ms duration using an

Agilent Technologies 8114A pulse generator and an avalanche photodiode (APD,

PerkinElmer SPCM-AQR-13) with an event counter (Nanoharp 250, Picoquant GmbH) as the detector.

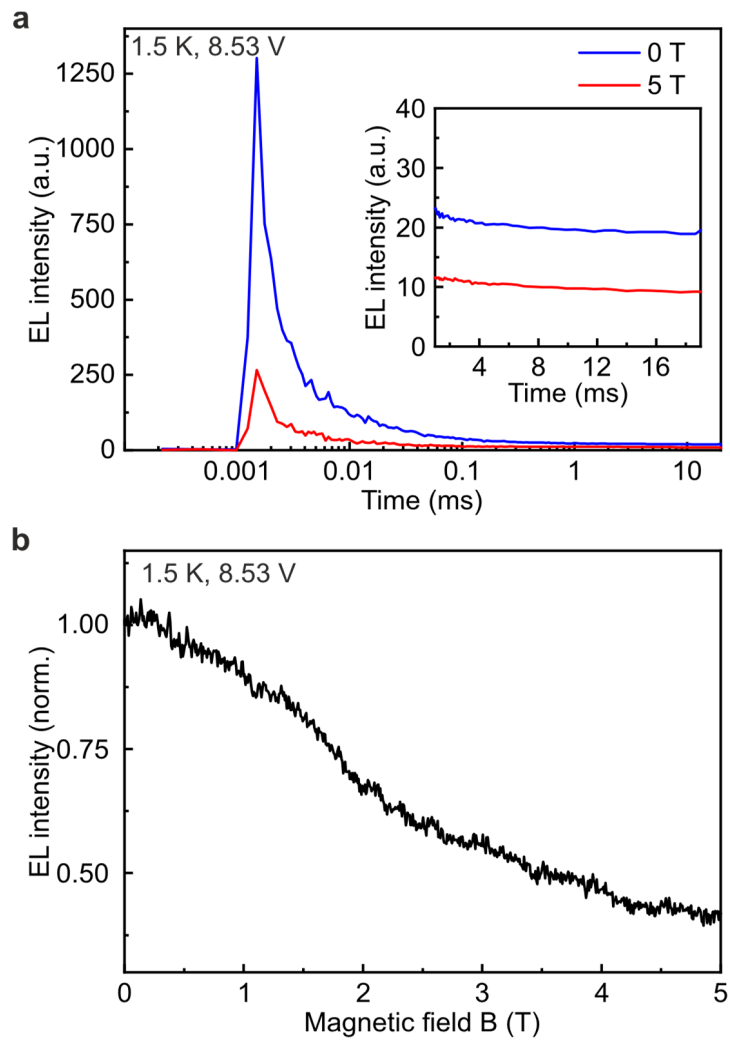

Figure S1. **Impact of TSP on the transient EL intensity at turn-on of a SYPPV OLED.** a) The EL intensity is dominated by an overshoot in the first 10 μs of the voltage pulse. In contrast to the report of Wang *et al.*,[1] we find no indication in the data of triplet-triplet annihilation occurring. The EL intensity is quenched immediately at the pulse onset by TSP at high fields, implying that spin relaxation takes place prior to electron-hole recombination. It is therefore not possible to extract information on the spin relaxation time from the transient data. After reaching equilibrium in the EL intensity, shown in the inset, the transient EL is suppressed by the magnetic field by the same amount as in the static MEL measurements at this voltage, shown in b).

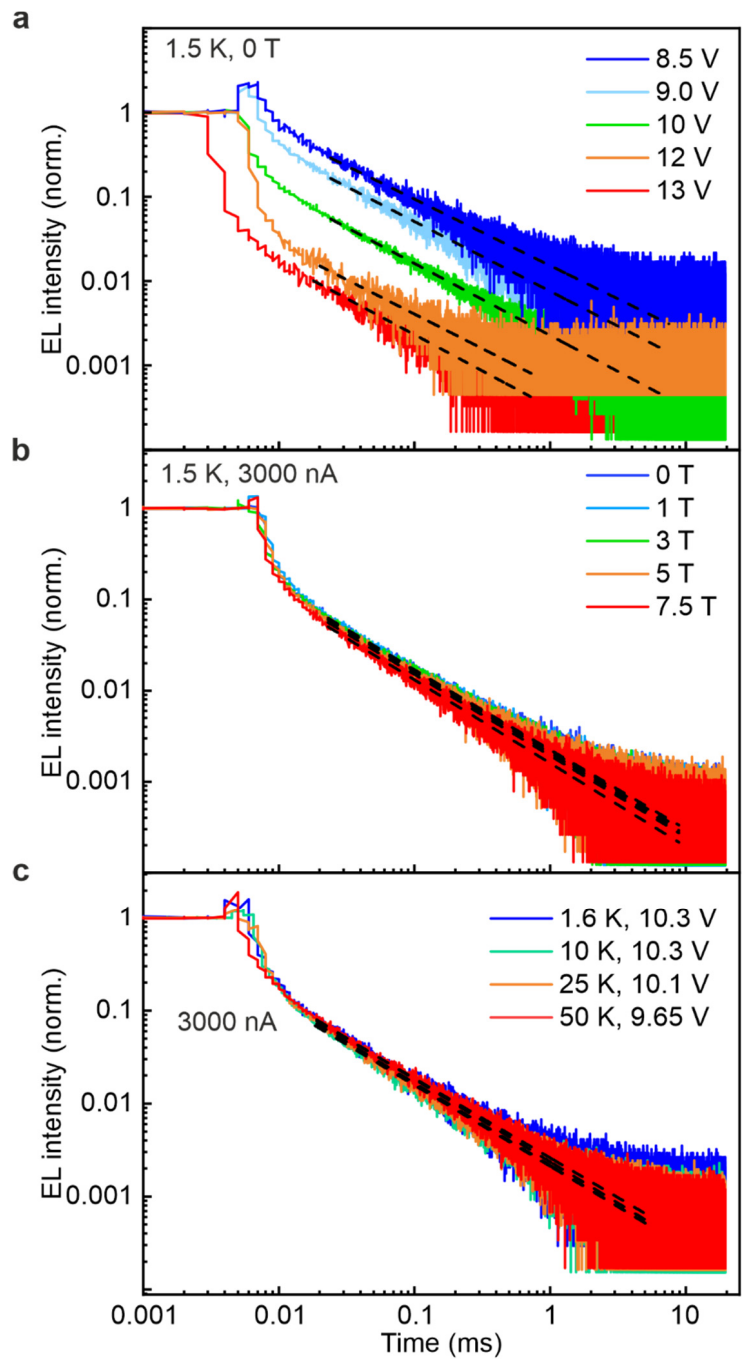

Figure S2. **Transient EL intensity of a SYPPV OLED at turn-off of a voltage pulse.** All data are normalized to the corresponding steady-state EL intensity. After a sharp EL overshoot, which is caused mainly by trapped charges, a power-law decay of the EL intensity is seen. Dashed lines indicate a power-law functionality of the form $t^{-\alpha}$, with α between 0.7 and 0.9.[2] a) Dependence of the decay dynamics on pulse voltage at zero field and 1.5 K. b) Dependence of the decay dynamics on magnetic field, following a pulse driving the OLED at a constant current of 3000 nA at 1.5 K. c) Dependence of the decay dynamics on temperature, following a pulse driving the OLED at a constant current of 3000 nA at zero field. Since the temperature slightly affects the resistance of the device, the pulse voltage is adjusted to reach the equilibrium current of 3000 nA. The dynamics are evidently insensitive to temperature and magnetic-field strength, implying that the transient EL at turn-off is not associated with delayed fluorescence from triplet-triplet annihilation but instead arises solely from trapped charges.

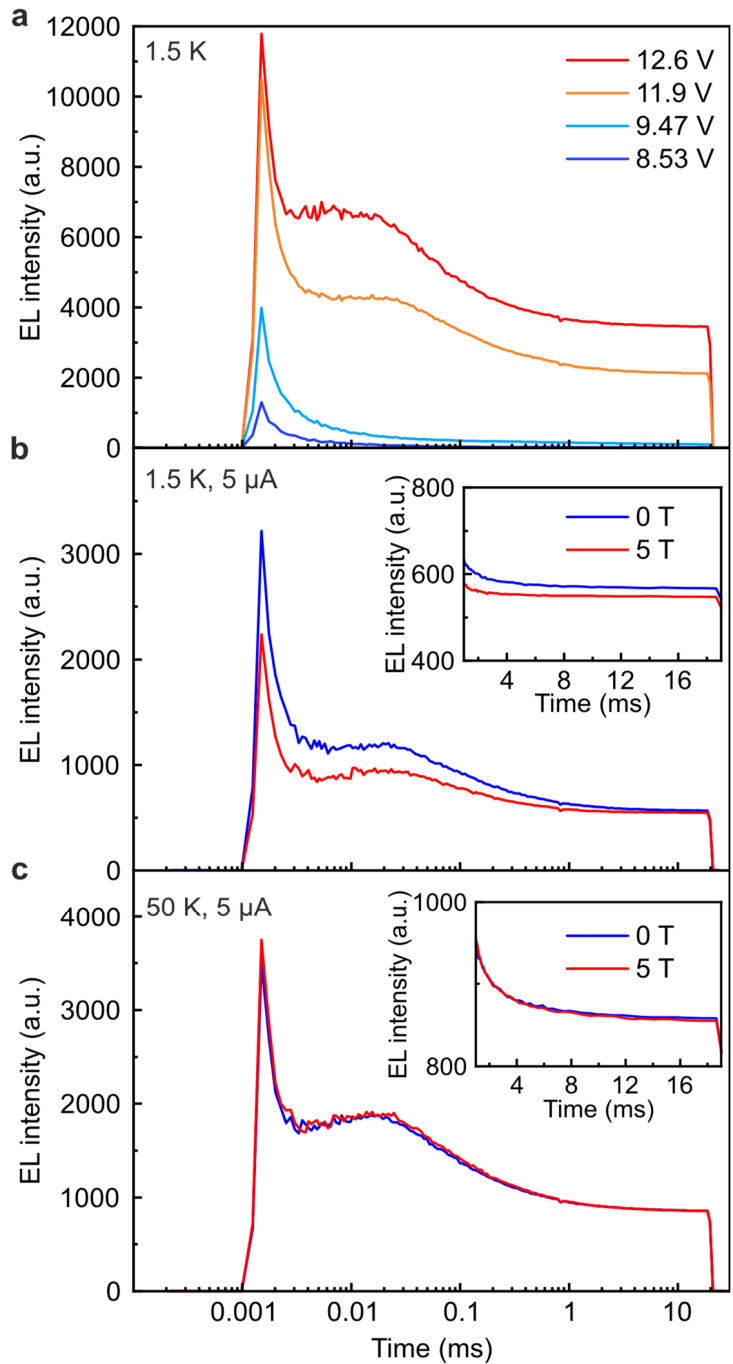

Figure S3. **Transient EL intensity of a SYPPV OLED over the entire voltage pulse of 20 ms duration.** a) Voltage dependence. At high voltages an additional plateau in the EL intensity is seen. The sharp initial EL overshoot appears to be reduced at higher voltages, and is thus presumably linked to charge accumulation at internal energy barriers within the device. After approximately 1 ms, steady-state conditions are reached. b) Magnetic-field dependence at 1.5 K. After reaching the steady-state EL intensity, illustrated in the inset, the EL is reduced

by a total of 5 % at 5 T due to TSP. However, suppression at this field of the initial overshoot and the plateau is much stronger. c) Magnetic-field dependence at 50 K. The magnetic field has no impact on the turn-on behaviour and on the steady-state EL intensity plotted in the inset, which is consistent with the absence of TSP at this temperature. The voltage of the pulse was chosen such that it corresponds to a steady-state device current of 5 μA.

**Fitting functions and parameters**

As noted in the main text, the thermal equilibrium value of spin-up and spin-down particles is determined by the Boltzmann distribution,

$$P_\uparrow^{\text{eq}} = \frac{1}{1 + \exp\left(\frac{g\mu_{\text{B}}B}{k_{\text{B}}T}\right)}$$

and

$$P_\downarrow^{\text{eq}} = 1 - P_\uparrow^{\text{eq}}, \tag{S1}$$

with $g$ the electron g-factor, $\mu_{\text{B}}$ the Bohr magneton and $k_B$ Boltzmann's constant. Taking into consideration a spin relaxation time $\tau_s$ and an effective free-carrier residence time $\tau_c$, the fraction of spin-polarised charge carriers at recombination is given by

$$P_\uparrow = \left(0.5 - P_\uparrow^{\text{eq}}\right) \exp\left(-\frac{\tau_c}{\tau_s}\right) + P_\uparrow^{\text{eq}}. \tag{S2}$$

With these considerations, one obtains the following formation probabilities of singlet and triplet states[1], namely

$$P_{\text{S}} = P_\uparrow \cdot P_\downarrow = P_\uparrow - P_\uparrow^2 \tag{S3}$$

and

$$P_{\text{T}} = 1 - P_{\text{S}}. \tag{S4}$$

For the pure MEL response due to TSP, the following fit functions were used

$$\text{MEL}_{\text{S}}^{\text{TSP}}(B) = \frac{P_{\text{S}}(B)}{P_{\text{S}}(0)} = 4 \cdot \left(P_\uparrow - P_\uparrow^2\right) \tag{S5}$$

$$\text{MEL}_\text{T}^\text{TSP}(B) = \frac{P_\text{T}(B)}{P_\text{T}(0)} = \frac{4}{3} \cdot (1 - P_\text{S}) \qquad (S6)$$

with the assumption that, at zero field, singlet and triplet states are formed in a ratio of 1:3. We note that this assumption is approximate because hyperfine fields can give rise to weak spin mixing, raising the singlet yield at zero field [3]. The only free fitting parameter is the ratio $\tau_\text{c}/\tau_\text{s}$.

To account for the $\Delta$g-mixing mechanism, a Lorentzian functionality of the form

$$\alpha \frac{\Delta B_{1/2}^2}{\Delta B_{1/2}^2 + B^2}$$

is added to the yields of singlet and triplet excitons, where $\alpha$ defines the relative strength of the effect and $\Delta B_{1/2}$ is the width of the Lorentzian that reflects the difference in precession frequencies of electron and hole within the pair. To describe the MEL response on a phenomenological level, the following expressions were used:

$$\text{MEL}_\text{S}(B) = \gamma_\text{S}\left(P_\uparrow - P_\uparrow^2\right) + \alpha_\text{S} \frac{\Delta B_{1/2}^2}{\Delta B_{1/2}^2 + B^2} + d_\text{S} \qquad (S7)$$

$$\text{MEL}_\text{T}(B) = \gamma_\text{T} \cdot \left(1 - P_\uparrow + P_\uparrow^2\right) - \alpha_\text{T} \frac{\Delta B_{1/2}^2}{\Delta B_{1/2}^2 + B^2} + d_\text{T}. \qquad (S8)$$

Here, $\gamma$ denotes the relative strength of TSP and $d$ accounts for a static offset correction. These two equations describe the measured MEL functionality at high magnetic fields well, but the parameters extracted should be viewed with caution because eq. (S7) and (S8) are merely phenomenological expressions. In particular, there is only limited physical meaning in the offset value and the fits are not strictly sufficiently constrained.

The fitted parameters for the MEL data shown in the main text are stated in the tables below.

Figure 2:

**panel a**

|  | 1.5 K | 2.5 K | 4.2 K | 10 K | 22 K |
|---|---|---|---|---|---|
| $\tau_c/\tau_s$ | 3.71 | 3.7 | 3.7 | 3.7 | 3.7 |

**panel b**

|  | 550 nA | 800 nA | 1000 nA | 3000 nA |
|---|---|---|---|---|
| $\tau_c/\tau_s$ | 5.56 | 2.1 | 1.00 | 0.37 |

**Table S1. Fit parameters for the SYPPV OLED dataset.** Parameters for temperature a) and current dependencies b) were obtained by fitting with eq. (S5).

Figure 3b):

**left**

|  | Fluorescence (S) | Phosphorescence (T) |
|---|---|---|
| $\tau_c/\tau_s$ | 4.01 | 4.01 |
| $d$ | 0.02 | -0.96 |

**right**

|  | Fluorescence (S) | Phosphorescence (T) |
|---|---|---|
| $\tau_c/\tau_s$ | 2.25 | 2.00 |
| $\Delta B_{1/2}$ (mT) | 3.80 | 3.72 |
| $\alpha$ | 0.22 | -0.31 |
| $\gamma$ | 4.172 | 0.350 |
| $d$ | -0.24 | 1.071 |

**Table S2. Fit parameters for Fig. 3.** Parameters in a) were obtained by using eq. (S5) and (S6). In b), the modified fit functions [eq. (S7) and (S8)] were used.

Figure 4

**panel a**

| Fluorescence (S) | 2 µA | 10 µA | 100 µA |
|---|---|---|---|
| $\tau_c/\tau_s$ | 1.02 | 0.741 | 0.499 |
| $\Delta B_{1/2}$ (mT) | 2.276 | 8.29 | 8.38 |
| $\alpha$ | 0.508 | 0.3661 | 0.38 |
| $\gamma$ | 0.2 | 3.037 | 2.637 |
| $d$ | 0.43 | -0.139 | -0.605 |

**panel b**

| Phosphorescence (T) | 2µA | 10 µA | 100 µA |
|---|---|---|---|
| $\tau_c/\tau_s$ | 1.045 | 0.715 | 0.501 |
| $\Delta B_{1/2}$ (mT) | 2.186 | 13.99 | 17.01 |
| $\alpha$ | -0.105 | 0.119 | 0.1005 |
| $\gamma$ | 0.1529 | 1.03 | 0.981 |
| $d$ | 1.016 | 0.362 | 0.363 |

**Table S3. Fit parameters for the MEL current dependency of DMDB in Fig. 4.**
Parameters were extracted by fitting eq. (S7) and (S8).

Figure 4

**panel c**

| Fluorescence (S) | 1.5 K | 5 K | 10 K |
|---|---|---|---|
| $\tau_c/\tau_s$ | 0.741 | 0.61 | 1.351 |
| $\Delta B_{1/2}$ (mT) | 8.29 | 7.974 | 9.94 |
| $\alpha$ | 0.3661 | 0.385 | 0.008 |
| $\gamma$ | 3.037 | 3.014 | 3.938 |
| $d$ | -0.139 | -0.158 | -0.004 |

**panel d**

| Phosphorescence (T) | 1.5 K | 5 K | 10 K |
|---|---|---|---|
| $\tau_c/\tau_s$ | 0.715 | 0.649 | 0.694 |
| $\Delta B_{1/2}$ (mT) | 13.99 | 3.9 | 4.69 |
| $\alpha$ | -0.119 | 0.036 | 0.0073 |
| $\gamma$ | 1.03 | 0.7234 | 1.208 |
| $d$ | 0.362 | 0.508 | 0.113 |

**Table S4. Fit parameters for the MEL temperature dependency of DMDB in Fig. 4.**
Parameters were extracted by fitting eq. (S7) and (S8).